\documentclass[aps,pra,reprint,a4paper,nofootinbib,superscriptaddress,numbers,longbibliography,showpacs,showkeys,floatfix]{revtex4-1}
\usepackage[utf8]{inputenc}
\usepackage[T1]{fontenc}
\usepackage{graphicx,amssymb,amstext,amsmath}
\usepackage{epsfig}
\usepackage[colorlinks=true, citecolor=blue, linkcolor=blue, urlcolor=blue, anchorcolor=blue]{hyperref}
\usepackage{epstopdf}
\usepackage[english]{babel}
\usepackage[autostyle]{csquotes}
\usepackage{lipsum}
\usepackage{subfigure,url}
\usepackage{color,graphics}
\usepackage{braket}
\usepackage{float}
\usepackage{mathtools}
\usepackage[dvipsnames]{xcolor}
\usepackage[pass,paperwidth=8.5in in,paperheight=11in]{geometry}
\usepackage{comment}
\draft

\begin{document}

\title{Revivals in One-dimensional Quantum Walks with a Time and Spin-dependent Phase Shift}

\author{Muhammad Sajid}
\email{m.sajid@kust.edu.pk}
\affiliation{Department of Physics, Kohat University of Science and Technology, Kohat 26000, Khyber-Pakhtunkhwa, Pakistan}

\author{Qurat ul Ain}

\author{Hanifa Qureshi}

\author{Tulva Tayyeba}
\affiliation{Government Girls Postgraduate College, Kohat, Khyber-Pakhtunkhwa, Pakistan}

%\date{\today}
\begin{abstract}
We investigate the role of a time and spin-dependent phase shift on the evolution of one-dimensional discrete-time quantum walks.
By employing Floquet engineering, a time and spin-dependent phase shift ($\phi$) is imprinted onto the walker's wave function while it shifts from one lattice site to another.
For a quantum walk driven by the standard protocol we show with our numerical simulations that complete revivals with equal periods occur in the probability distribution of the walk for rational values of the phase factor, i.e., $\phi/2\pi = p/q$.
For an irrational value of $\phi/2\pi$ our results show partial revivals in the probability distribution with unpredictable periods, and the walker remains localized in a small region of the lattice.
We further investigate revivals in a split-step quantum walk with a time and spin-dependent phase shift for rational values of $\phi/2\pi$.
In contrast to the case of standard protocol, the split-step quantum walk shows partial revivals in the probability distribution.
Furthermore, in view of an experimental realization we investigate the robustness of revivals against noise in the phase shift.
Our results show that signatures of revivals persist for smaller values of the noise parameter, while
revivals vanish for larger values. 
Our work is important in the context of quantum computation where quantum walks with a time and spin-dependent phase shift can be used to control and manipulate a desired quantum state on demand.
\end{abstract}

\maketitle

\section{Introduction}
Quantum walks (QWs) are dynamical tools that are used to control the motion of a quantum particle in space and time. Due to their potential applications in quantum information and physical sciences, they have been at the focus of much research work since its first introduction \cite{aharonov}. Owing to their quantum mechanical resources, i.e., quantum superposition, quantum interference, and entanglement, QWs hold the promise to develop new algorithms for computations on quantum computers \cite{deutsch1992rapid,divincenzo1995quantum,farhi1998quantum, kempe,ambainis2003quantum,computation1,computation2, chandrashekar2010discrete,venegas2012quantum}. In Physics QWs provide a versatile platform to simulate various physical phenomena, e.g., topological phases \cite{Kitagawa2010, Kitagawa2012, Kitagawa2012a, Asboth2012, Asboth2013, Asboth2014, Tarasinski2014, Asboth2015, Obuse2015, Cedzich2016BECorspnds, Groh2016, Xiao2017, Zhan2017, Sajid2019}, Anderson localization \cite{disorder1, Ahlbrecht2011, Ahlbrecht2011a, disorder2, disorder3}, Bloch Oscillations \cite{Genske2013, Cedzich2013, Arnault2020}, molecular binding \cite{Ahlbrecht2012, Lahini2012, Krapivsky2015}, and Hofstadter spectrum \cite{SajidThesis, Cedzich2020}, to name just a few. Due to their broad spectrum of applications, QWs have been realized in experiments using different physical systems, e.g., neutral atoms trapped in optical lattices \cite{karski, robens}, trapped ions on a line \cite{exp2, Xue2009, zahringer}, photons in free space \cite{broome, schreiber}, correlated photons on continuously evanescently coupled waveguides \cite{exp1}, and integrated photonics \cite{sansoni}.

QWs exhibit different features from their classical counterparts owing to their quantum mechanical resources.
In a quantum walk (QW) a quantum particle, which can exist in a superposition of several quantum states, moves to explore several paths simultaneously. 
Quantum interference takes place when different trajectories cross each other. As a result the probability distribution of a QW is strikingly different from a classical random walk. For example, the variance of a quantum walk grows quadratically faster with the number of steps of the walk compared to the linear growth of a classical random walk \cite{kempe,venegas2012quantum}.
QWs have shown speedups in comparison to their classical counterparts and, hence, are useful tools to design new fast algorithms for computation on a quantum computer \cite{ambainis2003quantum,search1}, and to simulate and control certain quantum computational tasks \cite{chandrashekar2010discrete}.
This has sparked a great interest to engineer different types of QWs and investigate their properties in different settings in the context of controlling and manipulating a desired quantum state for quantum computation and simulation.
In this vein, QWs have been investigated on a large scale which has resulted in the exploration of various types of walks. This includes QWs with decoherence \cite{decoherent,Romanelli2005,Kendon,Zhang2013,albertinDecoherence2014}, QWs with time-dependent coin \cite{Banuls2006,Xue2015,Cedzich2016,panahiyan,Katayama2020}, QWs with position dependent coin \cite{Wojcik2004,Uzma2020}, and QWs with different types of phase defects \cite{zhang2014one,Zhang2014TW0,Farooq2020}.

In this paper we engineer QW protocols (one standard QW protocol and one split-step protocol) which imprint a time and spin-dependent phase shift (TSDPS) to the wave function of a quantum particle undergoing a QW on a one-dimensional (1D) lattice. We get inspiration from the previous works \cite{Xue2015,Cedzich2016,panahiyan,Katayama2020,zhang2014one,Zhang2014TW0,Farooq2020,Cedzich2019} where a desired time evolution of a QW is achieved either by manipulating the coin parameter by making it time or position dependent, or by introducing different types of spatial phase defects into the wavefunction of the quantum particle. Our engineered QWs share common features, e.g., complete revivals and partial revivals in the probability distribution of QWs which are investigated in \cite{Xue2015,Cedzich2016,Katayama2020,Farooq2020}. We use the phase factor ($\phi$) as a control knob to control and manipulate the evolution of a QW. By numerically computing the probability distribution $P(x,n)$ and the standard deviation $\sigma(n)$ of the walk, and the probability of the particle to return to its initial position $P(x=x_{\text{i}},n)$ as a function of the number of steps of the walk, we show revivals in the evolution of QWs with TSDPS.
For rational values of the phase factor, i.e., $\phi/2\pi=p/q$, where $p$ and $q$ are mutually coprime integers, periodic revivals occur in the QW driven by the standard protocol where the period of revivals depends on the denominator $q$. In this case the quantum particle takes a finite excursion in the lattice and then comes back to its initial position. 
For an irrational $\phi/2\pi$, the QW shows partial revivals with irregular periods, and the walker remains localized in a small region of the lattice, hence, showing no transport. The periods of partial revivals in this case are not regular due to the incommensurability of the $\phi/2\pi$ with the step size of the QW. 
In the case of a QW walk driven by the split-step protocol, partial revivals occurs for rational values of $\phi/2\pi$ with periods of revivals double of the standard QW.

In experiments, imperfections are inevitable and, hence, imprinting an exact rational or irrational $\phi/2\pi$ is challenging. From this perspective we investigate the robustness of revivals against imperfections in the imprinted phase factor. We show with our numerical results that in the presence of a linear random noise (random fluctuations) in the imprinted phase factor, signatures of revivals persist for smaller values of the noise (fluctuations) parameter in the case of $\phi/2\pi=p/q$. Increasing the strength of the noise parameter results in suppression of revivals for both rational and irrational $\phi/2\pi$. In this case the quantum particle starts to spread out showing quantum transport.

The rest of the paper is organized as follows. In Sec.~\ref{QW_TSDPS} we introduce our system, the standard QW protocol, coin operator, and shift operator with TSDPS. We present our numerical results for revivals and partial revivals in the evolution of the QW driven by the standard protocol for rational and irrational values of $\phi/2\pi$, respectively. In Sec.~\ref{SPLIT_STEP} we introduce the split-step protocol with TSDPS, and investigate revivals in the probability distribution for rational values of $\phi/2\pi$. Section~\ref{QW_TSDPS_Noise} presents the effects of random fluctuations in the phase factor on revivals in the probability distribution of the QW driven by the standard protocol. We summarize our results and conclude with a brief outlook in Sec.~\ref{conc}.

\section{Quantum Walks with a Time and Spin-dependent Phase Shift} \label{QW_TSDPS}
We consider a single quantum particle (also called a walker) with two internal degrees of freedom undergoing a QW on a 1D lattice. The internal states of the particle (also called spin states due to its analogy with a spinor particle) are represented by basis vectors $\{\ket{s}:\ s\in \{\uparrow, \downarrow\}\}$ which span a two-dimensional Hilbert space $\mathcal{H}^s$. The position states of the walker are represented by the lattice coordinates $ x$ with basis vectors $\{ \ket{x}:  x \in \mathbb{Z} \}$ spanning a 1D Hilbert space $\mathcal{H}^x$. The quantum particle undergoing a quantum walk resides in a Hilbert space that is the tensor product of the two Hilbert spaces, i.e., $\mathcal{H}^s \otimes \mathcal{H}^x$. For simplicity, we use dimensionless units by assuming the lattice constant and the time duration of a single-step of the QW to be equal to 1.

The 1D standard protocol of the QW with TSDPS consists of a set of unitary operators. We call this protocol as the ``walk operator'' which is defined as,
\begin{align}
\label{eq:_1}
\hat{W}_{\phi}(n) = \hat{S}_x(\phi n) \, \hat{C},
\end{align}
here $\hat{C}$ is known as a coin operator 
and $\hat{S}_x(\phi n)$ as a spin-dependent shift operator. The coin operator acts on the internal states of the walker and rotates its spin state in the two-dimensional Hilbert space $\mathcal{H}^s$. In this work we employ the so-called Hadamard coin which is defined as,
\begin{align}
\label{eq:_coin}
\hat{C} = \frac{1}{\sqrt{2}} \begin{pmatrix} 1 & \ \ \ \ 1 \\ 1 & \ \ -1 \end{pmatrix} \otimes \ket{x}\bra{x}.
\end{align}
The spin-dependent shift operator $\hat{S}_x(\phi n)$ translates the walker by one lattice site to the right or to the left depending on its internal state, and imprints a step and spin-dependent phase shift $\phi$ to the walker's wavefunction. The shift operator is defined as,
\begin{align} \label{eq:shift-x}
      \hat{S}_x (\phi n) = & \sum_{x} 
           \Big[ \exp\big[i \phi n\big] \ket{\uparrow}\bra{\uparrow} \otimes \ket{x+1}\bra{x} \nonumber \\ & 
                + \exp\big[ - i \phi n\big] \ket{\downarrow}\bra{\downarrow} \otimes \ket{x-1}\bra{x} \Big],
\end{align} 
which shifts the walker in the spin-up (spin-down) state to the right (left). By applying the walk operator ( given in eq. (\ref{eq:_1})) to an initial state $(\ket{\Psi_{\text{i}}})$ of the walker one time constitutes a single step of the QW.
The evolution of the walk results by applying the walk operator to the initial state of the walker repeatedly for a large number of times. After certain $n$ number of steps of the walk (where $n\in \mathbb{N}$) the final state of the walker can be written as,
\begin{equation} \label{eq:final-state}
\ket{\Psi_n(\phi)}=\hat{W}_{\phi}(n) \ket{\Psi_{\text{i}}}.
\end{equation}
For a fixed value of $\phi$ the factor $n$ ensures the time dependence of the phase factor that is imprinted to the wavefunction of the walker at each step of the walk. The spatial probability distribution $P(x,n)$ is obtained by tracing out the coin degrees of freedom, i.e.,
\begin{equation} \label{eq:prob-dist}
P(x,n)=\sum_{s=0,1} \braket{\Psi_n(\phi)|\Psi_n(\phi)}.
\end{equation}
For rational values of the phase factor, i.e., $\phi/2\pi=p/q$, the walk operator $\hat{W}_{\phi}(n)$ is periodic, i.e., $\hat{W}_{\phi}(n+q*r) = \hat{W}_{\phi}(n)$ for some $r\in\mathbb{N}$. As a result the evolution of the walk (and hence, $P(x,n)$) is also periodic \cite{Cedzich2016}, i.e., $P(x,n+q*r) =P(x,n)$. To investigate revivals in the evolution of the QW we compute the return probability $P(x=x_{\text{i}},n)$ of the walker to its initial position $x_{\text{i}}$, and the standard deviation $\sigma(n)$ of the walk. After certain $n$ number of steps of the QW, $P(x=x_{\text{i}},n)$ and $\sigma(n)$ are obtained using the following expressions,
\begin{equation} \label{eq:prob-init-state}
P(x=x_{\text{i}},n)=\sum_{s=0,1} \braket{\Psi_{\text{i}}|\Psi_n(\phi)},
\end{equation}
\begin{equation} \label{eq:standard-deviation}
\sigma(n)= \sqrt{<x^2> - <x>^2}.
\end{equation}
\begin{figure*}[t]
  \centering
      \includegraphics[width=170mm]{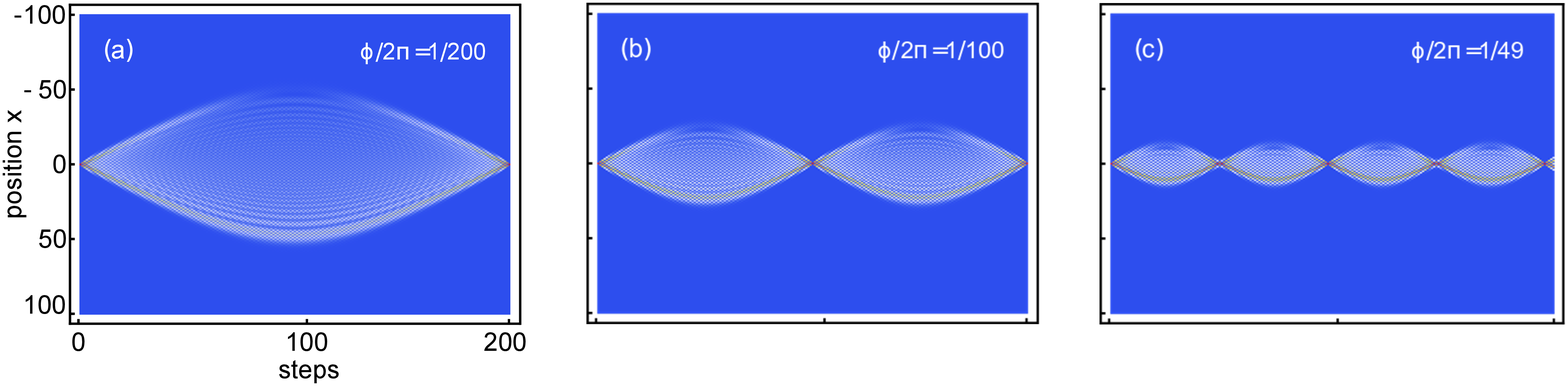} \\
      \includegraphics[width=170mm]{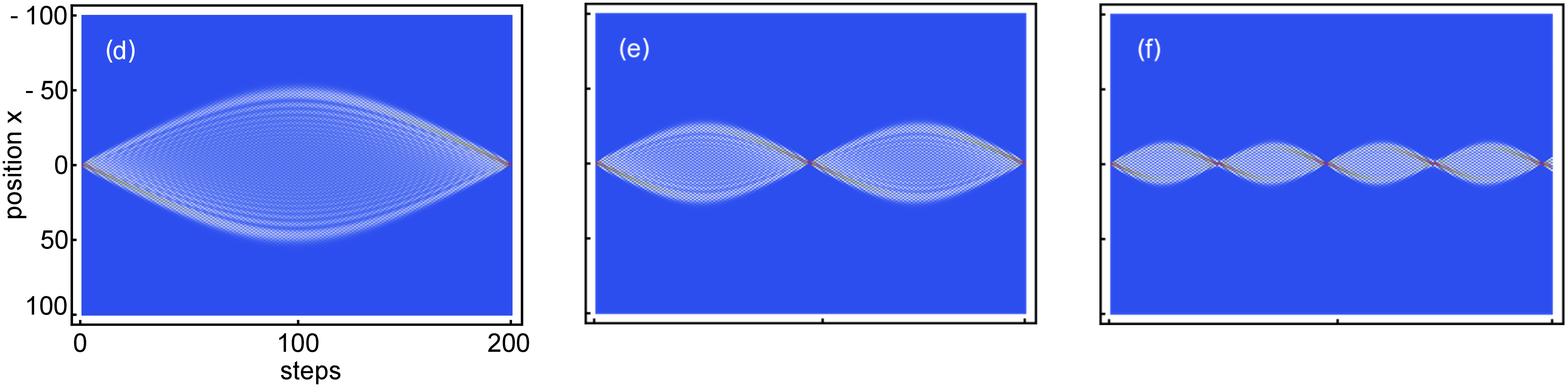} \\
      \includegraphics[width=170mm]{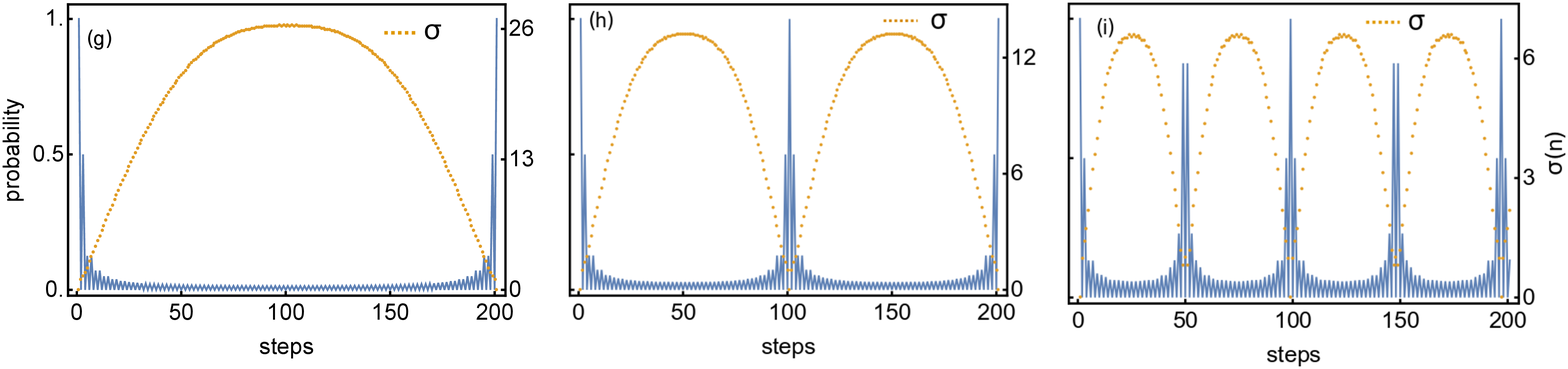}
%\center
\caption{(First two rows). Probability distribution of the QW with TSDPS in the position-time plane for rational values of $\phi/2\pi$ (red, white, and blue colors indicate the maximum ($1.0$), moderate ($\sim0.1$), and minimum ($\sim0.$) probabilities, respectively).
First row: The evolution of the walk with a quantum particle initially prepared in the initial state $\Psi_{i1}$. The values of $\phi/2\pi$ are indicated in the inset.
Second row: The evolution of the walk with a quantum particle initially prepared in the initial state $\Psi_{i2}$. The values of $\phi/2\pi$ remain same in a given column.
The difference in the probability distribution for the two initial states $\Psi_{i1}$ and $\Psi_{i2}$ can be noticed by a careful inspection.
Third row: The evolution of the return probability of the walker to its initial position (solid curve), and the standard deviation (dotted curve) of the walk upto $200$ steps of the walk. 
The behaviors of $P(x=x_{\text{i}}, n)$ and $\sigma(n)$ are same for both $\Psi_{i1}$ and $\Psi_{i2}$, and hence, the results are shown only for the first choice of the initial state $\Psi_{i1}$.
(a) For $\phi/2\pi=1/200$ and (b) $\phi/2\pi=1/100$ complete revivals occur after $n=q=200$ and $n=q=100$ steps of the walk, respectively. The return probability is equal to $1$ as it can be clearly seen from the corresponding subfigures (g) and (h) in the third row. (c) For $\phi/2\pi=1/49$ there occurs a partial revival when $n$ is an odd integral multiple of $q=49$ (i.e., $P(x=x_{\text{i}}, n)<1$), and a complete revival occurs when $n$ is an even integral multiple of $q=49$. The occurrence of revivals, their periodicities, and the boundedness of the standard deviation are independent of the initial state of the walker.}
\label{fig:a}
\end{figure*}
Here $<x>$ represents the average or expected value of the walker position $x$. Like $P(x,n)$, for $\phi/2\pi=p/q$ the return probability and the standard deviation of the walk are also periodic, i.e., $P(x=x_{\text{i}},n+q*r)$=$P(x=x_{\text{i}},n)$ and $\sigma(n+q*r)=\sigma(n)$. If initially a walker spreads from its initial position $x=x_{\text{i}}$, and after certain $n$ number of steps of the walk the return probability $P(x=x_{\text{i}}, n)$ becomes equal to $1$ (correspondingly $\sigma(n)$ becomes zero) we call this a complete revival. Similarly, if the walker returns to its initial position $x=x_{\text{i}}$ but with $P(x=x_{\text{i}},n)<1$ (correspondingly $\sigma(n)>0$), we call it a partial revival.
In the following sections we will investigate revivals in the probability distribution of a QW with TSDPS driven by the standard protocol (given in Eq.~(\ref{eq:_1})), and in a QW with TSDPS driven by the split-step protocol (given in Eq.~(\ref{eq:_SS_W_operator})) for various values of $\phi/2\pi$.

\subsection{Standard protocol with TSDPS for rational phase factor} \label{QW_TSDPS_Rational}
In this section we investigate the evolution of a QW driven by the standard protocol Eq.~(\ref{eq:_1}) for rational values of the phase factors, i.e., $\phi/2\pi=p/q$, where $p$ and $q$ are mutually coprime integers. 
We consider a walker initially residing at the origin of a 1D lattice, and consider two possibilities for the internal state of the walker, i.e., an equal superposition of the two spin states, and a spin up state only. The two initial states of the walker can be written as,
\begin{align} \label{eq:initial-state}
\begin{aligned}
\ket{\Psi_{i1}} &=\frac{1}{2} \Big( \ket{\uparrow} + i \ket{\downarrow} \Big) \otimes \ket{0},  \\
\ket{\Psi_{i2}} & = \ket{\uparrow} \otimes \ket{0}.
\end{aligned}
\end{align}
We evolve these initial states by periodically applying the walk operator given in Eq.~(\ref{eq:_1}) for a large number of steps. To demonstrate revivals in the walk we calculate the probability distribution ($P(x,n)$) of the walk in the position-time plane, the probability of the walker to return to its initial position ($P(x=x_{\text{i}}, n)$), and the standard deviation ($\sigma(n)$) of the walk using Eqs.~(\ref{eq:prob-dist}), (\ref{eq:prob-init-state}), and (\ref{eq:standard-deviation}), respectively.

In Fig.~\ref{fig:a}, we show our numerically simulated results of the evolution of $P(x,n)$ in the position-time plane for $n=200$ steps of the walk for $\phi/2\pi=1/200$ (Fig.~\ref{fig:a}(a)), $\phi/2\pi=1/100$ (Fig.~\ref{fig:a}(b)), and $\phi/2\pi=1/49$ (Fig.~\ref{fig:a}(c)). The evolution shown in Figs.~\ref{fig:a}(a)-(c) is computed with the initial state $\psi_{i1}$ as defined in Eq.~\ref{eq:initial-state}. Revivals in the probability distribution are apparent in all three cases with period of revival equal to $n=q$ for even values of $q$ (Fig.~\ref{fig:a}(a) and (b)) and $n=2q$ for odd values of $q$ ((Fig.~\ref{fig:a}(c)). The difference in the periods of the revivals for the even and odd values of $q$ is due to the fact that after odd number of steps of the walk the probability of the walker at its initial position is zero \cite{Cedzich2016}. In the case of odd $q$ the probability distribution shows partial revivals when $n$ is an odd multiple of $q$, and complete revivals when $n$ is an even multiple of $q$. In all three cases, i.e., Fig.~\ref{fig:a}(a)-(c), the walker initially takes an excursion in a finite region of the 1D lattice when $n<q/2$, then reverts its direction after $n=q/2$ to come back to its initial position. As a whole the walker remains bounded in a finite region of the 1D lattice during its evolution.

Figures~\ref{fig:a}(d)-(f) show the evolution of the walk for the same values of $\phi/2\pi$ as in Figs.~\ref{fig:a}(a)-(c) (the value of $\phi/2\pi$ remains same in a given column) but with the second choice of the initial state, i.e., $\psi_{i2}$ defined in Eq.~\ref{eq:initial-state}. From the evolution of the probability distribution it is apparent that the occurrence of revivals and their periodicity are same as in the case of the first initial state $\psi_{i1}$. However, the spatial profile of the probability distribution $P(x,n)$ for a given number of steps $n$ is different for the two initial states. In the case of $\psi_{i1}$ the probability remains higher in position states indexed by positive integers, i.e., $x>0$ (let us call it lower branch of the $P(x,n)$) throughout the evolution of the walk. In the case of $\psi_{i2}$ higher probabilities alternate between the upper ($x<0$) and the lower ($x>0$) branches of the probability distribution, i.e., $P(x,n)$ remains higher in a single branch for certain range of $n$ and then switches to the other branch.    

In order to give a clear demonstration of complete and partial revivals in the QW with TSDPS, in Figs.~\ref{fig:a}(g)-(i) we show the evolution of the return probabilities $P(x=x_{\text{i}}, n)$ of the walker to its initial position, and the standard deviation $\sigma(n)$ of the walk. Both the initial states $\psi_{i1}$ and $\psi_{i2}$ show identical behavior of $P(x=x_{\text{i}}, n)$ and $\sigma(n)$, we therefore show these results only for the first choice of the initial state, i.e., $\psi_{i1}$. The values of the phase factor remain same in a given column. For $\phi/2\pi=1/200$ and $\phi/2\pi=1/100$ complete revivals are clearly visible, i.e., $P(x=x_{\text{i}}, n)=1$ when $n$ is an integral multiple of $q$. In the case of $\phi/2\pi=1/49$, there are partial revivals when $n$ is an odd integral multiple of $q$ as $P(x=x_{\text{i}}, n)<1$, and complete revivals occur when $n$ is an even integral multiple of $q$ as $P(x=x_{\text{i}}, n)=1$. The extent of the region in which the walker remains bounded during its evolution can be estimated from the standard deviation as these are directly related. For a smaller value of the phase factor $\phi/2\pi$ (e.g., $\phi/2\pi=1/200$) the walker has a large excursion, i.e., $\sigma(n) \sim 26 $, and a smaller excursion ($\sigma(n) \sim 7 $) for a larger phase factor (e.g., $\phi/2\pi=1/49$).
In general, the TSDPS inhibits the ballistic expansion of the QW and confine the walker to a finite region of the lattice.
The extension of the bounded region decreases with the increase in $\phi/2\pi$. This behaviour is apparent from both the probability distribution and the standard deviation of the walk.
\begin{figure*}[t]
    \centering 
    \includegraphics[width=57mm]{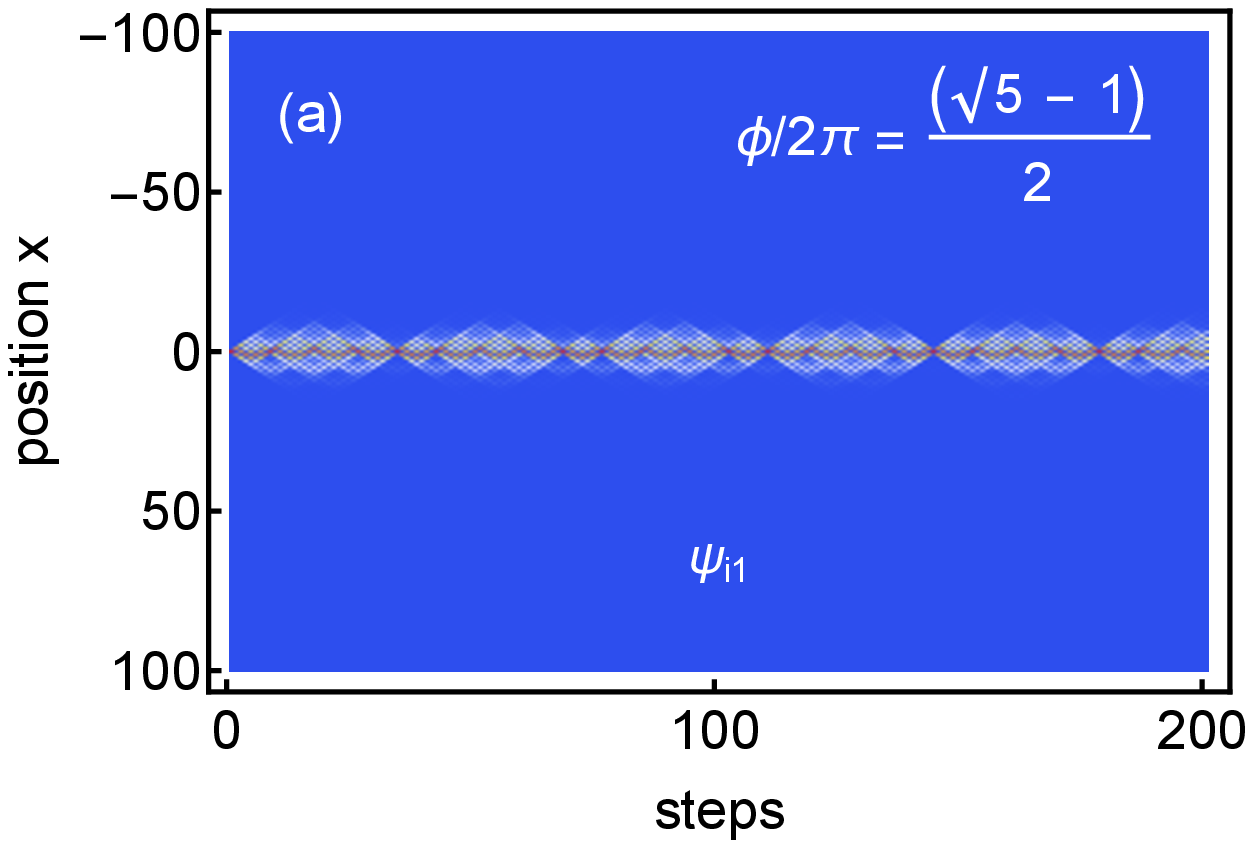}
    \includegraphics[width=57mm]{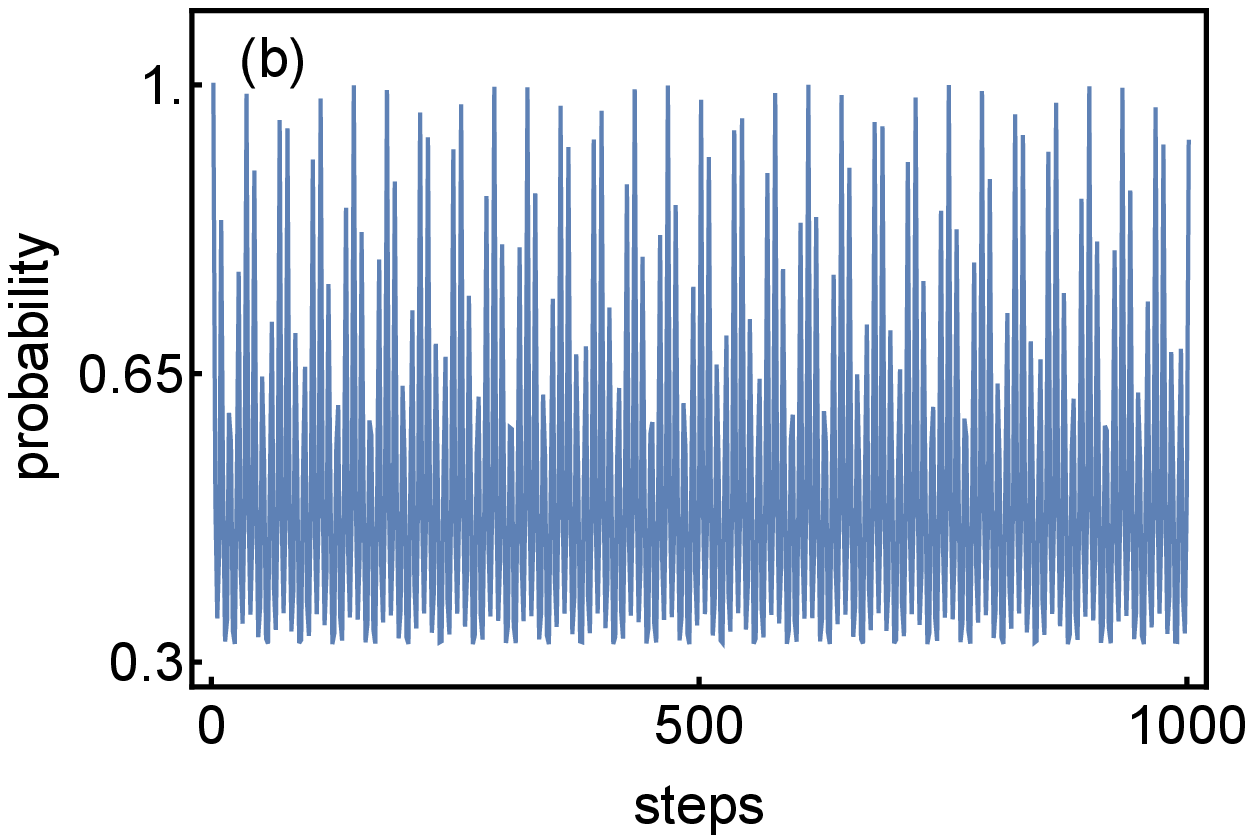} 
    \includegraphics[width=57mm]{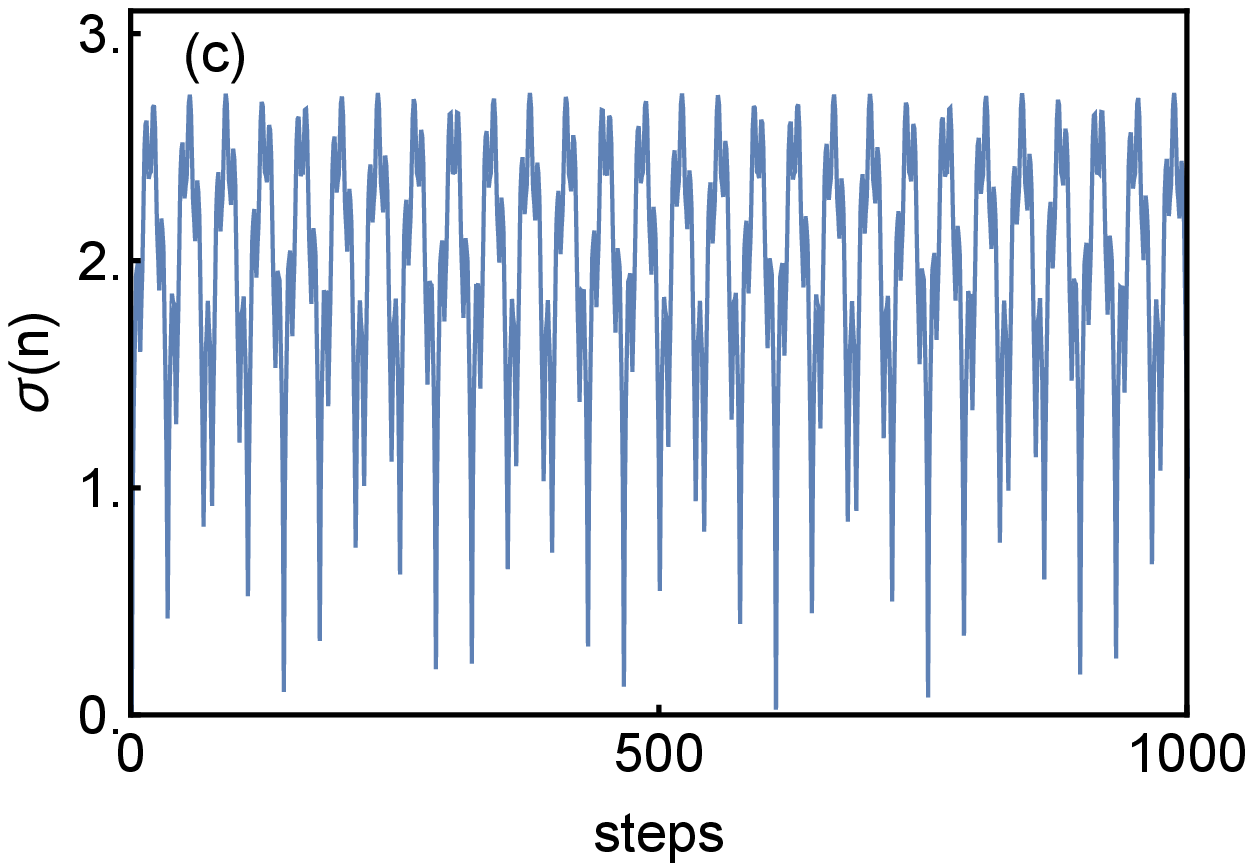} \\ 
	\includegraphics[width=57mm]{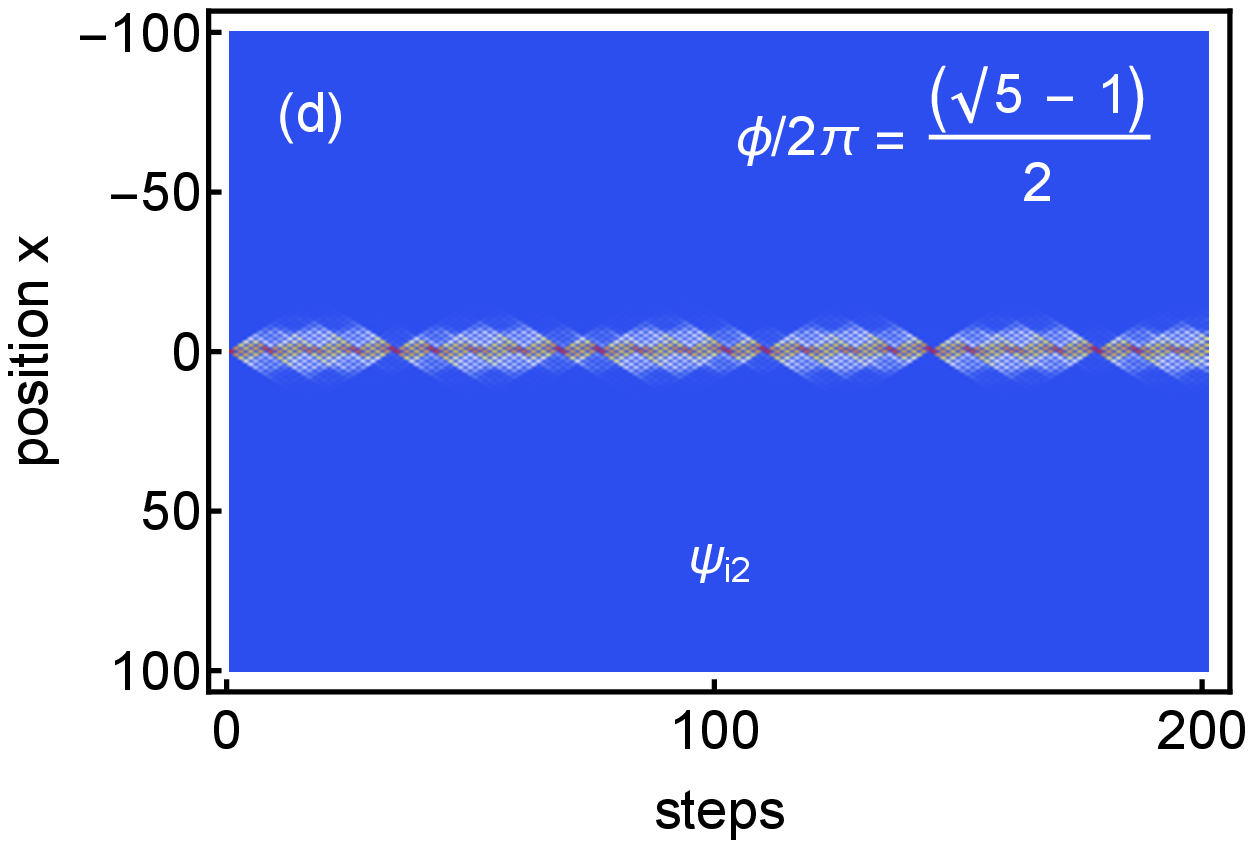} 
    \caption{Evolution of the QW with TSDPS and with the irrational phase factor $\phi/2\pi = (\sqrt{5}-1)/2$ (red, white, and blue colors indicate maximum ($1.$), moderate ($\sim0.1$), and minimum ($\sim0.$) probabilities, respectively). (a) and (d) show the probability distribution $P(x, n)$ of the walk in the position-time plane for $n=200$ and with symmetric (asymmetric) initial state $\Psi_{i1}$ ($\Psi_{i2}$) given in Eq.~(\ref{eq:initial-state}). (b) and (c) respectively show the long time evolution of the return probability $P(x=x_{\text{i}}, n)$ and the standard deviation of the walker with the initial state $\Psi_{i1}$ for $n=1000$. From (a) and (d) it is clear that the probability distribution is no more periodic for the irrational value of $\phi/2\pi$. The walker is completely localized in a finite region of the 1D lattice, which is also apparent from (c) which shows that $\sigma(n)$ has a finite upper bound, i.e., $\sigma<3$. The return probability in (b) shows a number of partial revivals with unpredictable periods in the long time evolution of the walk. $P(x=x_{\text{i}}, n)$ of the walker is bounded away from 0, i.e., $P(x=x_{\text{i}}, n) \ge 0.32$. 
    By carefully inspecting (a) and (d) one can notice that the probability distributions for the two choices of the initial states are not exactly similar. However, the return probability and the standard deviation show similar behavior for the two choices of the initial states. We have, therefore, shown $P(x=x_{\text{i}}, n)$ and $\sigma(n)$ for the first choice of the initial state, i.e., $\Psi_{i1}$. }
\label{fig:b}
\end{figure*}
\subsection{Standard protocol with TSDPS for an irrational phase factor} \label{QW_TSDPS_Irrational}
We now consider the case of an irrational phase factor $\phi/2\pi$. The well known approximate irrational number is the Golden ratio $(\sqrt{5} - 1)/2$. The walk operator (given in Eq.~(\ref{eq:_1}) with the phase factor equal to the Golden ratio is no more periodic as $(\sqrt{5} - 1)/2$ is incommensurate with the number of steps of the walk. However, the behaviour of the probability distribution of the walker still shows a number of partial revivals in its long time evolution. In Fig.~\ref{fig:b} we show the evolution of the probability distribution $P(x,n)$ for $n=200$ for both choices of the initial states ($\Psi_{i1}$ and $\Psi_{i2}$), the evolution of the return probability of the walker with initial state $\Psi_{i1}$ to its initial position $P(x=x_{\text{i}}, n)$ for $n=1000$, and the standard deviation $\sigma(n)$ of the walker with initial state $\Psi_{i1}$ for $n=1000$. Figure~\ref{fig:b}(a) shows that the walker remains completely localized in a small region of the lattice throughout its evolution. Figure~\ref{fig:b}(b) shows a number of revivals with unpredictable periods in the long time evolution of the walk. These revivals are not strictly complete as the return probability $P(x=x_{\text{i}}, n)$ is not exactly equal to unity for all $n>1$. Note that $P(x=x_{\text{i}}, n)$ is bounded away from $0$, i.e., $P(x=x_{\text{i}}, n) \ge 0.32$ for all $n$ showing that there is a significant probability of the walker to remain at its initial position during the evolution of the walk. The reason for this behavior is that the initial state of the walker has a significant overlap with a bound state \cite{Cedzich2013}. Figure~\ref{fig:b}(c) shows the evolution of the standard deviation of the walk which has an upper bound, i.e., $\sigma(n)<3$ for all values of $n$. This shows that the walker takes a very short excursion during the evolution of the walk and remains localized in a small region of the 1D lattice. Figure~\ref{fig:b}(d) shows the same evolution of the walk as in the Fig.~\ref{fig:b}(a) but with the second choice of the initial state, i.e., $\Psi_{i2}$. By a careful inspection of Fig.~\ref{fig:b}(a) and Fig.~\ref{fig:b}(d) one can notice that the probability distributions for the two initial states are not similar. However, the return probability, the standard deviation, and the localization behaviour of the walk are same for the two initial states. We, therefore, show the return probability and the standard deviation for the first initial state, i.e., $\Psi_{i1}$ only. 
\section{Split-Step Protocol with TSDPS for a rational phase factor} \label{SPLIT_STEP}
We now investigate revivals in a QW with a TSDPS driven by a split-step protocol. We consider only rational values of the phase factor, i.e., $\phi/2\pi=p/q$. The walk operator for this protocol is defined as,
\begin{align}
\label{eq:_SS_W_operator}
\hat{W}_{ss, \phi}(n) = \hat{S}_{x}^{\downarrow}(\phi n) \, \hat{C}_2 \, \hat{S}_{x}^{\uparrow}(\phi n) \, \hat{C}_1,
\end{align}
\begin{figure*}[t]
    \centering 
    \includegraphics[width=71mm]{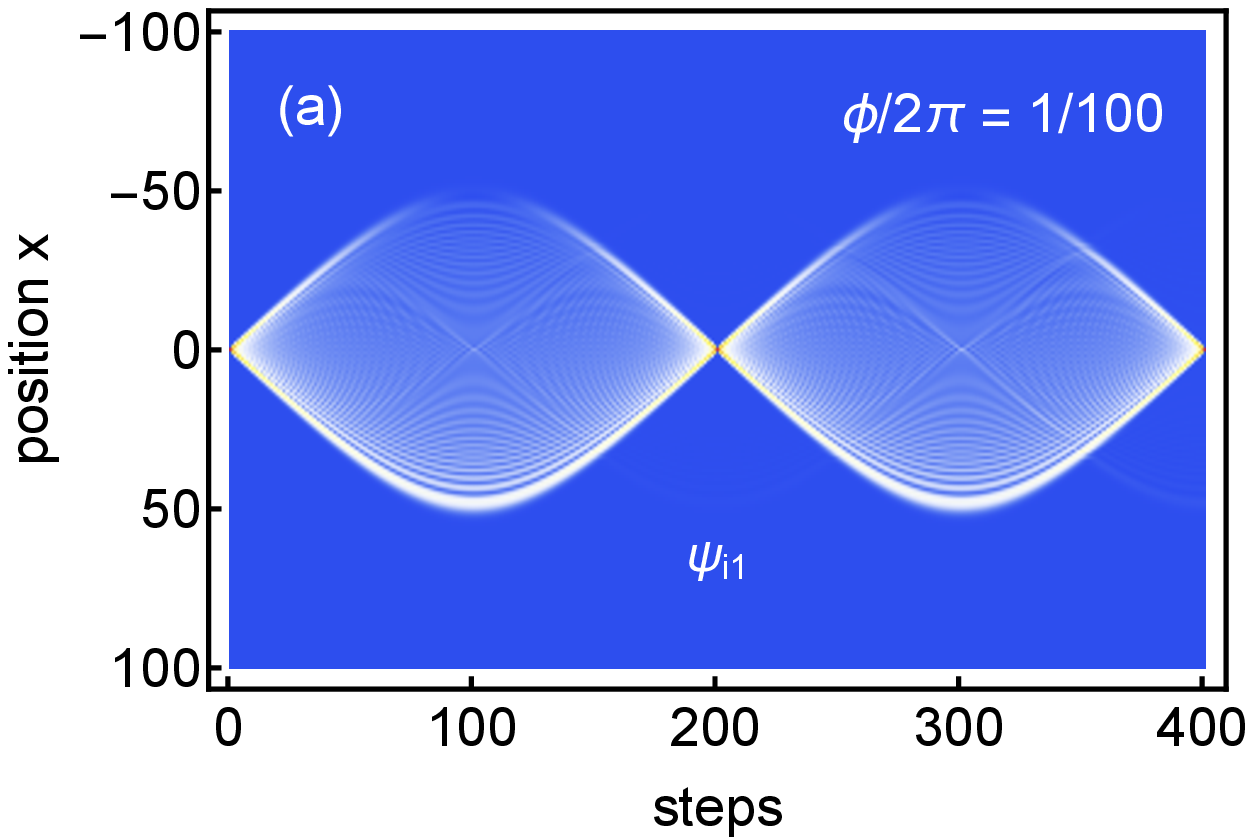}
    \includegraphics[width=75mm]{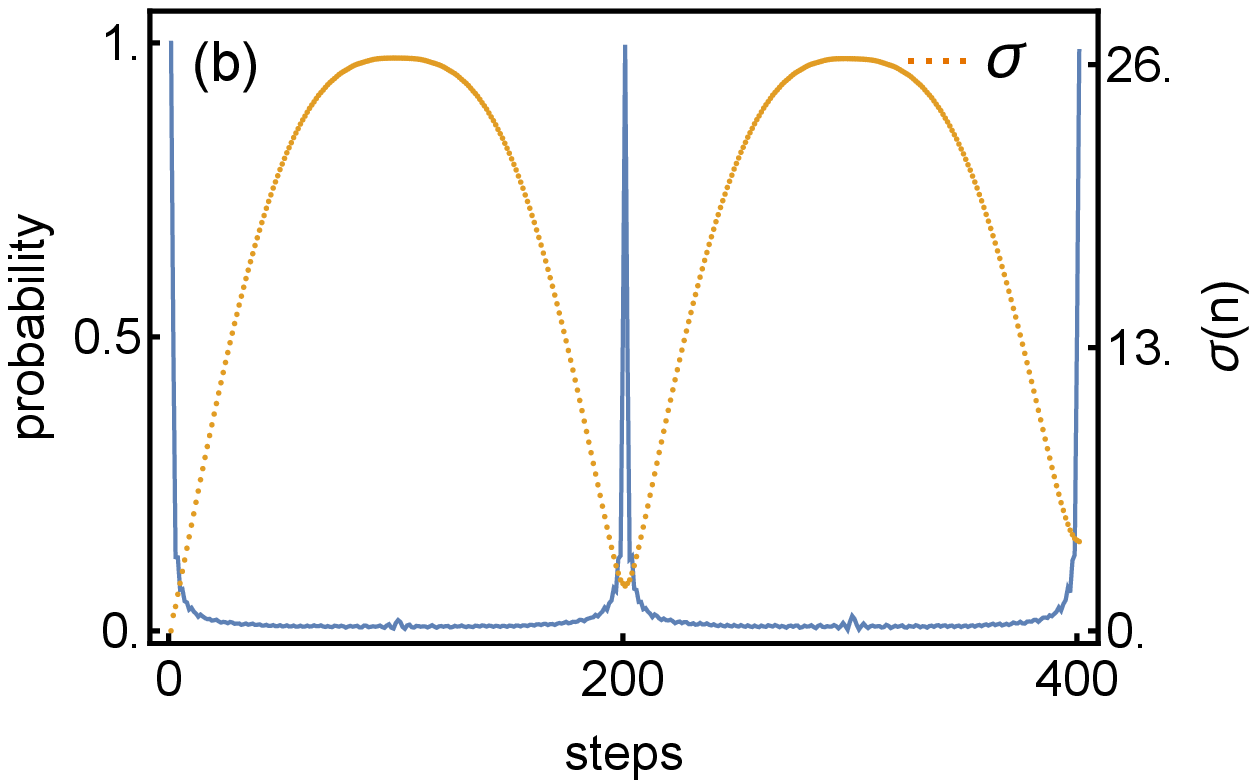} \\
    \includegraphics[width=71mm]{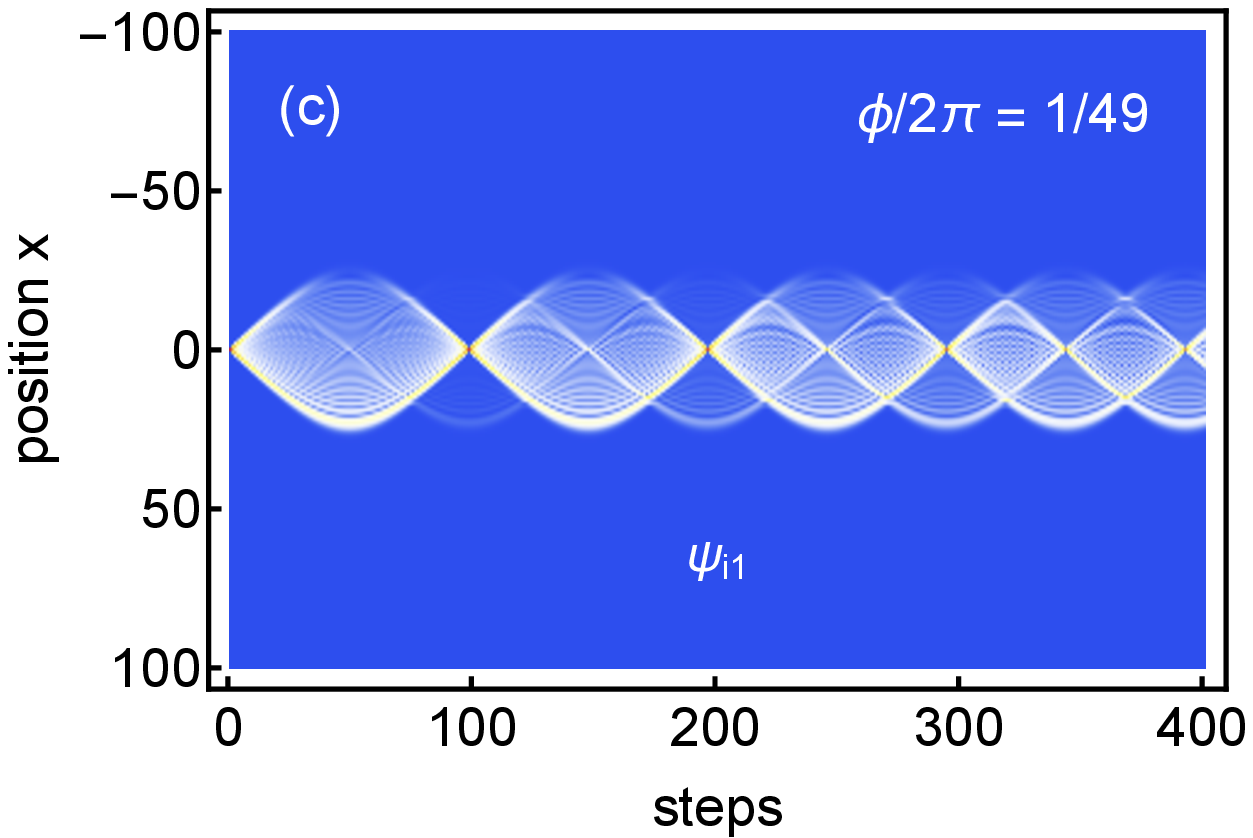}
    \includegraphics[width=75mm]{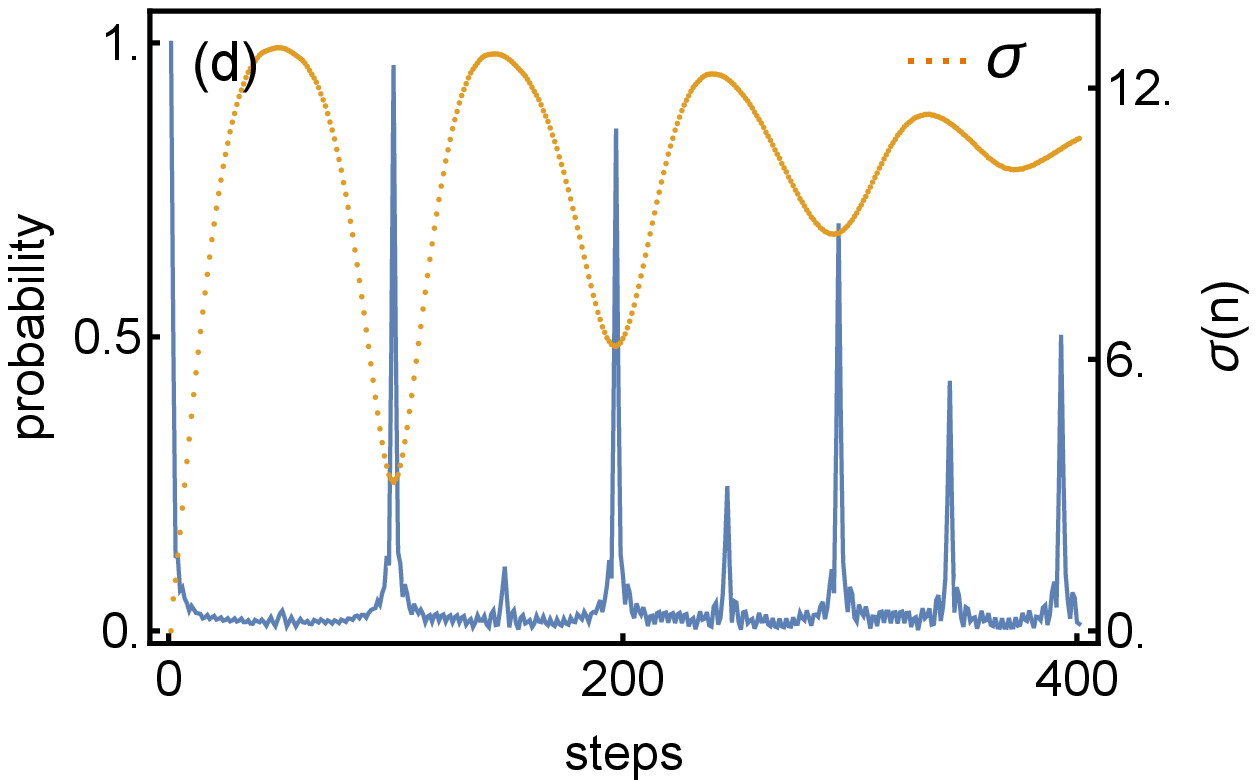}
\caption{Evolution of the split-step QW with TSDPS with rational phase factor $\phi/2\pi$ (red, white, and blue colors indicate maximum ($1.$), moderate ($\sim 0.1$), and minimum ($\sim 0.$) probabilities, respectively). (a) and (c) show the Probability distribution $P(x, n)$ of the walk in the position-time plane for $\phi/2\pi=1/100$ and $\phi/2\pi=1/49$, respectively.  (b) and (d) show the evolution of the return probability $P(x=x_{\text{i}}, n)$ (indicated by the blue solid curve) and $\sigma(n)$ (indicated by the orange dotted curve) corresponding to (a) and (c), respectively. 
It is clear from the probability distribution that the walker remains bounded in a finite region of the lattice, as it was the case in the QW with TSDPS driven by the standard protocol. The return probability remains smaller than unity in both (b) and (d), showing partial revivals. The periods of the partial revivals are double of the standard protocol case (compare with Fig.~(\ref{fig:a})). At each period of revivals the return probability decreases and the standard deviation increases with the number of steps of the walk. The standard deviation has an upper bound due to the fact that the walker remains bounded during the evolution of the walk.}
\label{fig:Split_Step}
\end{figure*}
which consists of two coin operators $\hat{C}_1$ and $\hat{C}_2$, and two shift operators $\hat{S}_{x}^{\uparrow}(\phi n)$ and $\hat{S}_{x}^{\downarrow}(\phi n)$. We consider both the coin operators to be the Hadamard coins as defined in Eq.~(\ref{eq:_coin}).
The shift operator $\hat{S}_{x}^{\uparrow} (\phi n)$ shifts only the spin-up state of the walker to the right by a unit length and imprints a TSDPS to its wavefunction leaving the spin-down state unchanged. Mathematically, it is defined as,
\begin{align} \label{eq:shift-x-up}
     \hat{S}_{x}^{\uparrow} (\phi n) =  & \sum_{x}
             \Big[ \exp\big[i \phi n\big] \ket{\uparrow}\bra{\uparrow} \otimes \ket{x+1}\bra{x} \nonumber \\ & 
              + \ket{\downarrow}\bra{\downarrow} \otimes \ket{x}\bra{x}\Big].
\end{align} 
Similarly, the operator $\hat{S}_{x}^{\downarrow}(\phi n)$ shifts only the spin-down state of the walker to the left by a unit length and imprints a TSDPS to its wavefunction leaving the spin-up state unchanged. It is defined as,
\begin{align} \label{eq:shift-x-up}
     \hat{S}_{x}^{\downarrow} (\phi n) =  & \sum_{x}
             \Big[ \ket{\uparrow}\bra{\uparrow} \otimes \ket{x}\bra{x} \nonumber \\ & 
              + \exp\big[- i \phi n\big] \ket{\downarrow}\bra{\downarrow} \otimes \ket{x-1}\bra{x}\Big].
\end{align} 
To investigate revivals in the QW driven by the split-step protocol with TSDPS, we consider a walker in an initial state $\Psi_{i1}$ defined in Eq.~(\ref{eq:initial-state}), and evolve it through the walk operator defined in Eq.~(\ref{eq:_SS_W_operator}) for a large number of steps. We numerically compute the probability distribution $P(x,n)$, the return probability $P(x=x_{\text{i}},n)$ to the initial position, and the standard deviation $\sigma(n)$ of the walker similar to the QW driven by the standard protocol.

In Fig.~(\ref{fig:Split_Step}) we show our numerical results for $P(x,n)$, $P(x=x_{\text{i}},n)$, and $\sigma(n)$ of the split-step QW with TSDPS. The evolution is carried out with the initial state $\Psi_{i1}$ for $n=400$ steps of the walk. Figures~\ref{fig:Split_Step}(a) and (c) show $P(x,n)$ in the position-time plane for $\phi/2\pi=1/100$ and $\phi/2\pi=1/49$, respectively. The probability distributions show periodic behavior. The walker spreads initially for $n<q$, then reverts its direction around $n=q$, and comes back at its initial position $x=x_{\text{i}}$ after $n=2q$. However, the return probability $P(x=x_{\text{i}},n)$ remains smaller than unity for both even and odd $q$. This is clearly visible in Figs.~\ref{fig:Split_Step}(b) and (d) which show partial revivals in the probability distribution. The periods of revivals are double ($n=2q$ for even $q$) of the QW driven by the standard protocol (compare these results with Fig.~(\ref{fig:a}) for the corresponding values of $\phi/2\pi$). For an odd value of $q$ the period of revival is $n=2q$ which is same as it was in the case of QW with TSDPS driven by the standard protocol.

Figures.~\ref{fig:Split_Step}(b) and (d) show that at a given period of a partial revival the value of $P(x=x_{\text{i}},n)$ decreases compared to its value at a previous revival. This shows that revivals will be faint after a few periods. However, the walker will still be bounded in a finite region of the lattice.
The standard deviation $\sigma(n)$ also shows the occurrence of partial revivals as it shows quasi-periodic behaviour. At a given period of a partial revival the value of $\sigma(n)$ is larger compared to its value at a previous revival. However, it remains bounded throughout the evolution of the walk.
In general, in the split-step QW with TSDPS the walker remains bounded in a finite region of the 1D lattice. The extent of the bounded region varies inversely to the phase factor $\phi/2\pi$ similar to the standard protocol case.
\begin{figure*}[t]
  \centering
  \includegraphics[width=75mm]{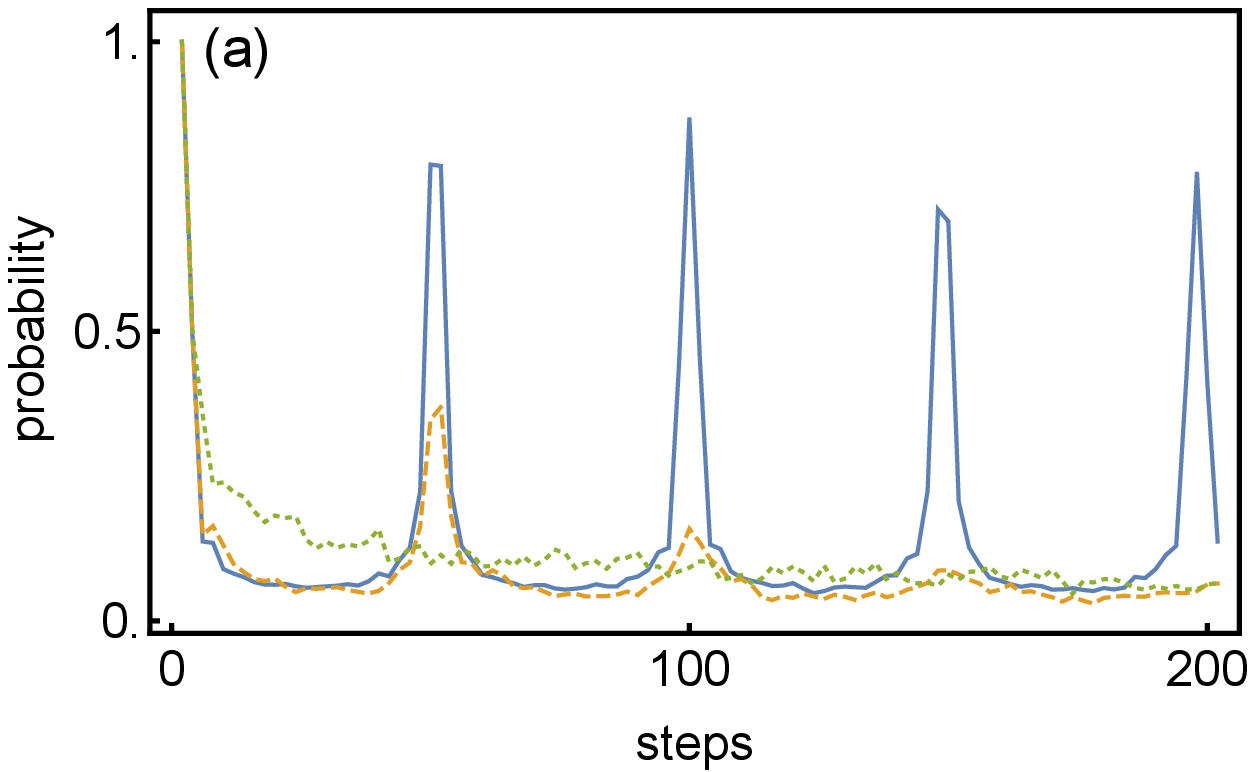}
  \includegraphics[width=75mm]{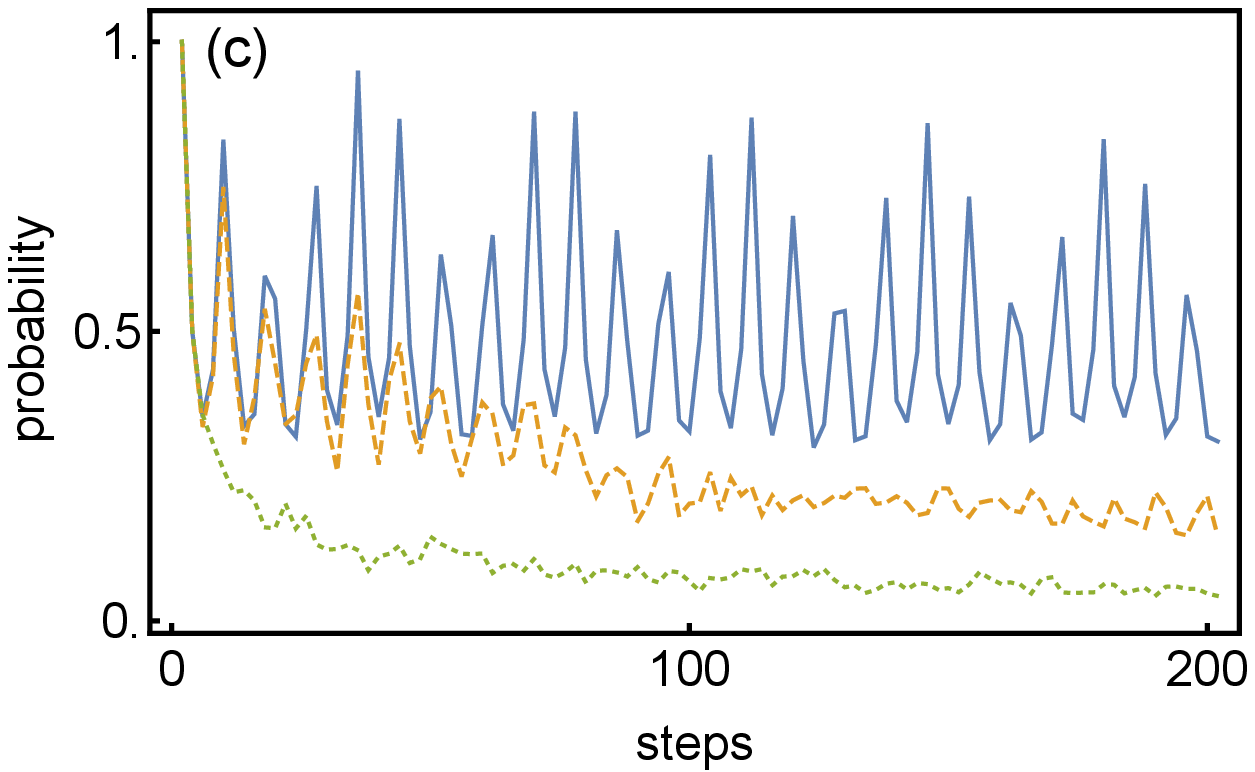}
      \\
  \includegraphics[width=75mm]{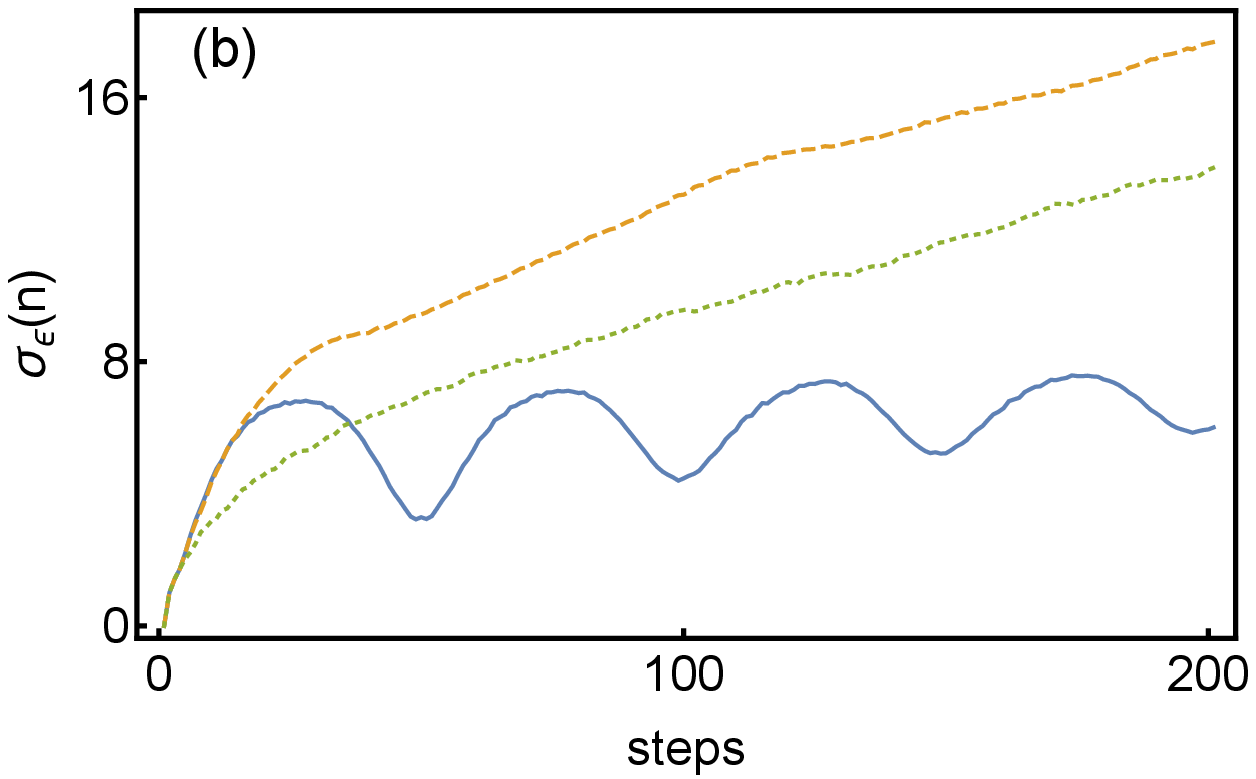} 
  \includegraphics[width=75mm]{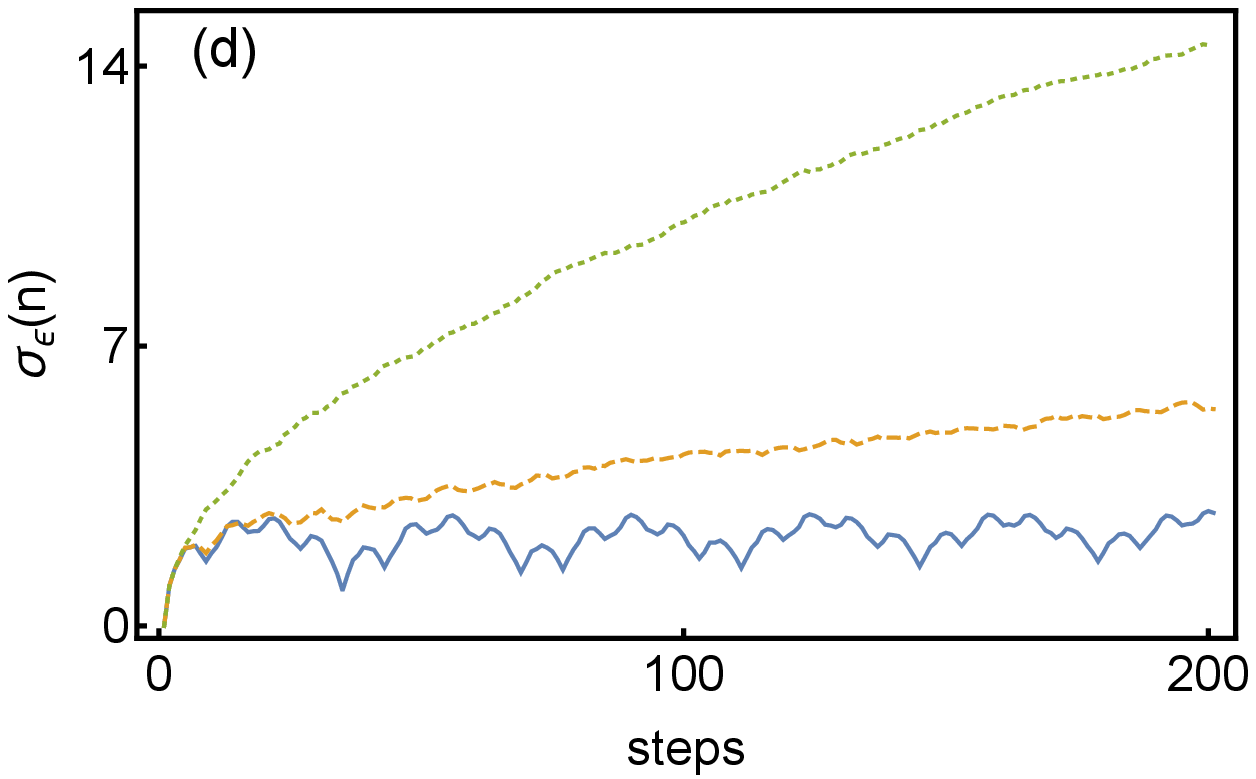} 
  \caption{Return probability $P_{\epsilon}(x=x_{\text{i}},n)$ and standard deviation $\sigma_{\epsilon}(n)$ of the QW driven by the standard protocol for rational and irrational values of $\phi$ with $\phi_{\epsilon}$ varying randomly from one step of the walk to another. (a) $P_{\epsilon}(x=x_{\text{i}},n)$, (b) $\sigma_{\epsilon}(n)$ for $\phi=2\pi/49$ and $\epsilon = 1/20$ (solid blue curve), $\epsilon = 1/5$ (orange dashed curve), and $\epsilon = 1$ (green dotted curve). Signatures of revivals are apparent for the first two values of $\epsilon$, and in the third case the walker starts diffusion and, hence, there are no revivals. (c) $P_{\epsilon}(x=x_{\text{i}},n)$, (d) $\sigma_{\epsilon}(n)$ for $\phi=2\pi(\sqrt{5} -1)/2$ and with the same three values of $\epsilon$ (with the same color scheme) as in (a) and (b). For $\epsilon = 1/20$ the return probability (represented by the solid blue curve) is bounded away from zero which is similar to the case discussed in \ref{fig:b}(b). By increasing $\epsilon$, i.e., $\epsilon=1$ the walker starts to spread and $P_{\epsilon}(x=x_{\text{i}},n)$ is no more bounded from below. $P_{\epsilon}(x=x_{\text{i}},n)$ approaches to zero for higher values of $n$. The results shown in this figure are average values of $20$ runs.} %
\label{fig:c}
\end{figure*}
\section{Effects of Noise in the Phase Factor} \label{QW_TSDPS_Noise}
In the QW driven by the standard protocol we have shown that exact revivals in $P(x=x_{\text{i}},n)$ occur for exact rational values of $\phi/2\pi$. However, imprinting an exact rational value of $\phi/2\pi$ in experiments will be challenging as imperfections are inevitable. This can possibly eradicate any sign of revivals in the probability distribution of the QW in experiments. It is therefore worth to investigate the effects of noise in the phase factor $\phi/2\pi$ on the revivals in the probability distribution. The purpose is to find how far the above studied revivals in $P(x_{\text{i}},n)$ are robust to fluctuations in the phase factor. 

We consider a phase factor $\phi_{\epsilon}(n)$ that is fluctuating randomly. Following \cite{Cedzich2016}, we model the fluctuating phase factor as,
\begin{equation} \label{eq:fluctuation_moeld}
\phi_{\epsilon}(n)=\phi + \epsilon \mathcal{R}_n,
\end{equation}
here, $\epsilon \in [0, 1]$ is the fluctuation parameter and $\mathcal{R}_n \in [-1, 1]$ is chosen randomly at each step of the walk. We investigate the evolution of the QW driven by the standard protocol with TSDPS that is fluctuating randomly from one step of the walk to another. Since we are interested to investigate the effects of random fluctuations in the phase factor on revivals, we compute only the return probability $P_{\epsilon}(x=x_{\text{i}},n)$ and the standard deviation $\sigma_{\epsilon}(n)$ of the QW. 

Figure.~\ref{fig:c} shows our numerical results of the evolution of the return probability $P_{\epsilon}(x=x_{\text{i}},n)$ and the standard deviation $\sigma_{\epsilon}(n)$ of the walker undergoing a QW with TSDPS subjected to random fluctuations in the phase factor. The results shown in this figure are averaged over $20$ runs. We consider both rational and irrational values of $\phi/2\pi$. Figures~\ref{fig:c}(a) and (b) show the effect of random fluctuations on revivals for $\phi=2\pi/49$. For $\epsilon=1/20$ (red solid line) and $\epsilon=1/5$ (blue solid line) the signatures of revivals persist, while for $\epsilon=1$ (black solid line) the walker starts to spread out and, hence, revivals are absent. This clearly indicates that by increasing the fluctuation parameter from certain limit, quantum transport takes place in the QW with TSDPS as the walker is no more localized.

Figure~\ref{fig:c}(c) shows the evolution of the return probability $P_{\epsilon}(x=x_{\text{i}},n)$ and Fig.~\ref{fig:c}(d) shows the standard deviation $\sigma_{\epsilon}(n)$ of the walk for $\phi=(\sqrt{5}-1)/2$ and $\epsilon=1/20$ (red solid line), $\epsilon=1/5$ (blue solid line), and $\epsilon=1$ (black solid line). For $\epsilon=1/20$ the return probability is bounded away from zero and the walker is completely localized. This is similar to the behavior shown in Fig.~\ref{fig:b}(b). However, for $\epsilon=1/5$ and $\epsilon=1$ the return probabilities are no more bounded from below as it decreases with $n$ and approaches to zero for larger $n$. This is in stark contrast to the exact irrational $\phi$ as the walker is no more localized and spreads out from its initial position.
\section{Conclusion}\label{conc}
We have studied the evolution of 1D QWs (standard protocol and split-step protocol) with time and spin-dependent phase shifts. For the QW with standard protocol we have investigated the cases of both rational and irrational phase factors $\phi/2\pi$. For rational  values, i.e., $\phi/2\pi=p/q$ the walk operator is periodic in time and our results show revivals in the evolution of the QW. The period of revivals is $n=q$ for even values of $q$ and $n=2q$ for odd values. For the irrational case the walk operator, and hence, the return probability is no more periodic. However, we found a number of partial revivals with unpredictable periods in the evolution of the walk. The walker is completely localized in a finite region of the lattice. 
For the QW with split-step protocol we considered only rational values of $\phi/2\pi$. We found partial revivals in the evolution of the walk with periods of partial revivals double (for even $q$) of the QW driven by the standard protocol.
We have shown that the results obtained for revivals, i.e., return probabilities and standard deviation do not depend on the choice of the initial state of the walker. However, the probability distributions are different for different initial states. 

From the point of view of an experimental realization of revivals in QWs with TSDPS driven by the standard protocol, we further investigated the effects of random fluctuations on revivals. For both rational and irrational cases of the phase factor, we found that signatures of revivals persist for smaller values of the noise parameter. By increasing the value of the fluctuation parameter from certain limit, signatures of revivals vanish and quantum transport takes place. This so-called noise-induced transport is demonstrated for both the rational and irrational values of the phase factor. Our study is important from the point of view that under which conditions one should expect revivals in QWs with TSDPS, how far it is robust against noise in the phase factor, and when quantum transport takes over revivals. A natural extension of this work is to investigate the role of TSDPS in QWs in higher dimensions. 
\section*{Acknowledgements}
Q. A., H. Q, and T. T acknowledge support from physics department, Kohat University of Science and Technology, Pakistan.

\twocolumngrid

\end{document}